\documentclass[footinbib,twocolumn,showpacs,amsmath,amstex,amssymb,mathfonts,superscriptaddress,prx]{revtex4}
\usepackage{graphicx}
\usepackage{color}
\usepackage{bm}

\usepackage{graphicx}
\usepackage{color}
\usepackage{bm}
\usepackage{amsmath}
\usepackage{amssymb}
\usepackage{amsthm}
\usepackage{amsfonts}
\usepackage{multirow}
\usepackage{makecell}
\usepackage{float}
\usepackage{comment}
\usepackage{enumitem}
\usepackage{verbatim}
\usepackage{mathtools}

\begin{document}

\title{Fermionic fractional quantum Hall states: A modern approach to systems with bulk-edge correspondence}
\author{Yoshiki Fukusumi and Bo Yang} 
\affiliation{Division of Physics and Applied Physics, Nanyang Technological University, Singapore 637371.}

\date{\today}
\begin{abstract}
In contemporary physics, especially in condensed matter physics, fermionic topological order and its protected edge modes are one of the most important objects. In this work, we propose a systematic construction of the cylinder partition corresponding to the fermionic fractional quantum Hall effect (FQHE) and a general mechanism for obtaining the candidates of the protected edge modes. In our construction, when the underlying conformal field theory has the $Z_{2}$ duality defects corresponding to the fermionic $Z_{2}$ electric particle, we show that the FQH partition function has a fermionic T duality. This duality is analogous to (hopefully the same as) the dualities in the dual resonance models, typically known as supersymmetry, and gives a renormalization group (RG) theoretic understanding of the topological phases. We also introduce a modern understanding of bulk topological degeneracies and topological entanglement entropy. This understanding is based on the traditional tunnel problem and the recent conjecture of correspondence between the bulk renormalization group flow and the boundary conformal field theory. Our formalism gives an intuitive and general understanding of the modern physics of the topologically ordered systems in the traditional language of RG and fermionization and may serve as a complement of more mathematical physical frameworks, such as fermionic category theories.
\end{abstract}

\maketitle 

\section{Introduction}
\label{sec:introduction}

Fermionic representations of topological orders (TOs) and critical systems are one of the most fundamental subjects in contemporary physics\cite{Kitaev:2006lla,Kitaev:2001kla,2014AnPhy.351.1026G,Shiozaki:2016zjg,Gaiotto:2015zta,Lan_2016}. For fermionic (e.g. electric) degrees of freedom in a lattice model, one can expect the emergence of the fermionic TOs and the corresponding topological quantum field theories (TQFTs).  Some of the celebrated examples can date back to the Abelian and non-Abelian bosonization in the high energy and condensed matter physics\cite{Witten:1983ar,Coleman:1974bu,Mandelstam:1975hb,Alvarez-Gaume:1986nqf,Affleck:1985wb,Haldane:1981zza}. It should also be noted that in the original works of the Goddard-Kent-Olive coset construction\cite{Goddard:1986ee,Goddard:1984hg} of the conformal field theories (CFTs), the Majorana and Symplectic fermionic representations of the minimal CFTs have already been introduced. However, a unified understanding of the fermionic TO and the criticality of its edge theory is still work in progress. Even when restricting our attention to the $(1+1)$ dimensional systems at criticality, systematic construction of the boundary and bulk  fermionic CFT is only accomplished very recently\cite{Hsieh:2020uwb,Fukusumi:2021zme,Weizmann,Kulp:2020iet,Runkel:2020zgg,Runkel:2022fzi}\footnote{As one of the authors has previously noted in the review part of \cite{Yao:2020dqx}, fermionic representations of CFT characters have been studied as Rogers-Ramanujun identity in the mathematical physics and integrable model communities \cite{Kedem:1992jv,Kedem:1993ze,Blondeau-Fournier:2017otv,Blondeau-Fournier:2010fkq,Schilling:1995fk}.}. It should be worth noting that the fermionic CFTs have a close relation to the boundary theory of the $(2+1)-$dimensional fermionic topological ordered systems under the bulk-edge correspondence. Related to this correspondence, the fermionic version of the categorical models such as the supermodular tensor category have been proposed\cite{Lan_2016,Aasen:2017ubm,Inamura:2022lun}. An important difference between a fermionic CFT and the corresponding bosonic CFT is that the former intrinsically contains nonlocal objects (e.g. the disorder operators). Typically, this results in the modular $T^{2}$ invariance of the torus partition function of the fermionic CFTs\cite{Ginsparg:1988ui}, which is closely related to the nonlocal nature of the ``gauging" operation\cite{Fukusumi_2022}.

\begin{figure}[htbp]
\begin{center}
\includegraphics[width=0.5\textwidth]{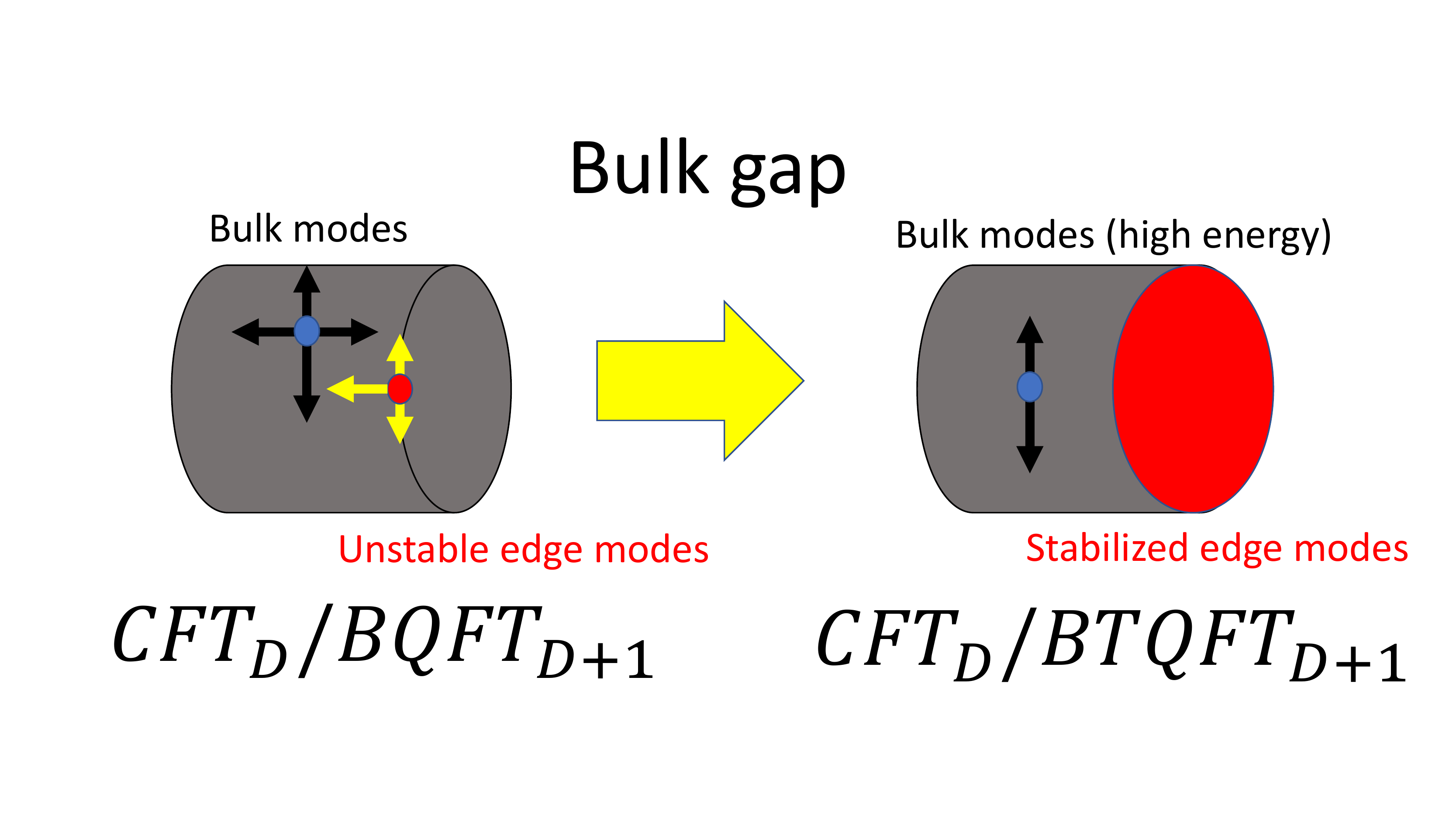}
\caption{Bulk and boundary RG picture of the bulk-edge correspondence. As we discuss in the subsequent sections, we expect the edge modes in $CFT_{D}/BTQFT_{D+1}$ as a (emergent) renormalizable theory.}
\label{QFT}
\end{center}
\end{figure}

An important connection between the TO in the $\left(2+1\right)-$dimension and the critical systems in the $\left(1+1\right)-$dimension is the bulk-edge correspondence, a systematic framework very useful for the construction and understanding of the topological phases of matter\cite{Schoutens:2015uia}. From a modern perspective, the bulk-edge correspondence is a salient feature for both the symmetry protected topological phases and the strongly correlated fractional quantum Hall effect where no symmetry protection is required\cite{Weerasinghe_2014,Jolicoeur:2014isa,Scaffidi_2016,Wen:2022lxh,Thorngren:2020wet}. In such systems, it is natural to understand the bulk-edge correspondence and the bulk gappability arguments as a kind of restricted bulk renormalization group flow (RG flow) starting from a gapless quantum field theory (QFT)\cite{2013PhRvX...3b1009L,Milovanovic:1996nj}. In the bulk and boundary RG flow perspective, one can summarize the procedures in the existing literature as the following:
\begin{enumerate}
\item Assuming the existence of gapless matter with bulk-edge correspondence, so that all such gapless excitations can be interpreted as either excitations at the edge, or quasihole (i.e. anyon) excitation in the bulk. Such bulk-edge correspondence can be thought of as a generalized version of the operator-state correspondence of massless QFT, thus the dispersion of these gapless excitations is generally assumed to be linear. One can expect this correspondence can be interpreted as the correspondence between the $CFT$ in $D-$dimension and the bulk quantum field theory in $D+1-$dimension (i.e. the $CFT_{D}/BQFT_{D+1}$ correspondence),  where $D$ is the space-time dimension.
\item Introduction of the bulk perturbations which preserve or break the bulk-edge correspondence. With this process, the correspondence can be reduced to the $CFT_{D}/BTQFT_{D+1}$ correspondence, and this results in the emergence of the edge modes (See FIG.\ref{QFT}).
\item Computing the boundary perturbations induced by the bulk RG flow. This aspect has captured attention only recently in \cite{Lichtman:2020nuw,Fukusumi:2020irh}, indicating the importance to study the effects of the boundary perturbations on the unstable boundary conditions under the bulk gap.
\end{enumerate}

The first point may be striking for the readers because we have invoked the bulk-edge correspondence which has been assumed in condensed matter literature implicitly as the RG flow of $CFT_{D}/BQFT_{D+1}$ correspondence. This view is a new generalization of the AdS/CFT correspondence and needs further verification. Our proposal opens up the possibility to formulate intuitive understanding of the correspondence in condensed matter in a more accessible way for different research fields.

The necessary and sufficient conditions for the bulk gappability in the second point for fermionic or more general TOs and their relations to the protected edge modes in FQHE are still under development\cite{2013PhRvX...3b1009L,Wang:2012am,Lan:2014uaa,Heinrich_2018,Kobayashi:2022vgz,Kaidi:2021gbs}. Moreover the third point, which should be treated as the defining property that protects the gaplessness of the edge modes, has rarely been discussed in the traditional RG way. In the existing literature, as far as we know, there exists no systematic analysis for the conformal dimensions of the perturbations of the edge CFT to determine whether the resulting edge modes are protected or not. The protectedness of the edge modes may be understood as a consequence of the bulk and boundary RG flow which typically leads to boundary duality or symmetry induced by the bulk anyons\cite{Lichtman:2020nuw,Fukusumi:2020irh}. This emergent symmetry or duality at the edge may suppress boundary irrelevant perturbations that break the conformal structure of the boundary theory\footnote{In the existing literature, it seems to be assumed that all perturbation automatically vanishes by the gapping operation to the bulk in such protected boundary conditions, but this condition might be difficult to realize in the traditional boundary RG framework. Hence the finite size scaling analysis for both bulk and boundary is necessary to establish the validity of the theoretical understandings of the protected edge modes.}. Remarkably, this `emergent' renormalizability is similar to the asymptotic freedom in high energy physics, and we expect it is useful for high energy physicists working on gauge-gravity duality\cite{Alday:2009aq}, for example. Related approach can be seen in \cite{Santachiara:2010bt,Kimura:2022zsx}.

It is also worth noting that whereas there exist several proposals to study the boundary degree of freedom in general BCFT, the protected edge modes appearing after the bulk and boundary RG flow can be linear combinations of the Cardy states\cite{Fukusumi:2020irh,Verresen:2019igf,Graham:2003nc}. Hence to formulate and test the protectedness of the edge modes, one needs to develop the truncated conformal space approach (TCSA) \cite{Yurov:1989yu,Yurov:1991my} to CFT in general space-time dimensions, with the bulk and boundary perturbations for such unstable boundary conditions. This is generally a difficult task, as the application of TCSA to systems with boundaries is limited to the lower dimensional CFTs\cite{Watts:2011cr,Feverati:2006ni,James_2018,Horvath:2022zwx} and that of higher dimensional CFT (without boundaries) is still under development \cite{Rychkov:2014eea,Rychkov:2015vap,Hogervorst:2014rta}. Moreover, even in these works, the boundary states  beyond the Cardy states have been rarely discussed. Related to this problem, we would also like to note recent proposals to attach bulk QFT to nontrivial lower dimensional CFT\cite{DiPietro:2019hqe,Behan:2020nsf,Behan:2021tcn}. These works imply nontrivial relations of the bulk and boundary theories outside of the usual $CFT_{D}/BTQFT_{D+1}$ correspondence.

In this work, we propose a systematic way to construct the fermionic partition functions corresponding to the edge modes of the fermionic FQHEs on a 2+1 dimensional cylinder starting from the bulk gapless quasihole excitations. Our proposal gives a unified and modern understanding of the appearance of the edge modes, whereas it is analogous to the traditional conformal bootstrap and S-matrix theory of dual models\cite{Veneziano:1968yb,Mandelstam:1974fq}\footnote{In FQHE, one couples CFT to the $U(1)$ model, instead of the gravity in the dual models.}. There exist several  proposals which are related to the present work, by using the Lagrangian formalism \cite{Ino:2000wa,Ino:2000ve,Gromov:2019cgu,Salgado-Rebolledo:2021nfj,Ma:2021dua,Nguyen:2022khy} or AdS/CFT correspondence\cite{Fujita:2009kw,Wu:2014dha,Mezzalira:2015vzn,Melnikov:2012tb,Ryu:2010fe,Bayntun:2010nx,Karch:2010mn}, but our formalism seems to be more elementary and closer to the original operator formalism of the dual models. Especially, we observe the fermionic T duality in the FQHE partition function when there exists the $Z_{2}$ duality defect in the underlying CFT\cite{Gliozzi:1976qd}. In this formalism, one can identify the Neveu-Schwartz (NS) sector of the cylinder partition function as the edge modes and the Ramond (R) sector as the bulk excitations, and this allows us to give a RG interpretation under the bulk gapping process\footnote{For the definition of NS and R sector in the spin systems, see discussions in\cite{Hsieh:2020uwb,Fukusumi:2021zme}. Roughly speaking, the NS sector corresponds to the periodic boundary conditions and the R sector corresponds to the anti-periodic boundary conditions.}. Our work is based on the recent findings of the fermionic CFT and BCFT\cite{Hsieh:2020uwb,Fukusumi:2021zme,Weizmann,Kulp:2020iet,Runkel:2020zgg,Runkel:2022fzi}. Related discussions in the condensed matter physics can be seen in the study of the correspondence between the Jack polynomial and the plane partition function of FQH states \cite{Bernevig_2008,Bernevig2008PropertiesON,Regnault2009TopologicalEA}, but our work is applicable to more general FQH states which can be constructed from the CFTs with $Z_{2}$ symmetry. 

Moreover, we propose a systematic understanding of the bulk topological degeneracies by combining the recent conjecture by Cardy\cite{Cardy:2017ufe,Lencses:2018paa} that relates the BCFT to the RG flow of CFT, with the tunnel problem that maps the closed manifold to the open manifold\cite{Wen:1990se}. Hence by applying the handle decomposition, our analysis leads to a general way of calculating the topological property of the $\left(2+1\right)-$dimensional system constructed from the $2-$dimensional CFT in arbitrary geometry. We also expect our analysis can be generalized to other space-time dimensions and this may give a new perspective on the CFT/TQFT correspondence \cite{Chen:2022wvy,Vanhove:2018wlb,Fuchs:2002cm,Fuchs:2004dz}. It should be stressed again that one should start the construction from the bulk gapless excitations with the bulk-edge correspondence, and this formulation can give a clearer understanding of a topological phase in the sense of RG.

The rest of the paper is organized as follows: In Sec.\ref{Laughlin}, we revisit the construction of the partition functions of the Laughlin state with emphasis on their modular property. This will be followed by the later sections where we emphasize the modular noninvariance of the half-flux quantum sectors of the bosonic partition function emerging from the coupling to the CFTs. In Sec.\ref{cylinder_fermion}, a construction of the cylinder partition function of the FQH state from the general fermionic CFT is shown. The partition function corresponding to $CFT_{D}/BQFT_{D+1}$ and its flow to $CFT_{D}/BTQFT_{D+1}$ is introduced. We also show the relation between the bosonic Kramers-Wannier $Z_{2}$ duality and the fermionic T duality and their implication from a modern perspective. In Sec.\ref{Ishibashi}, we discuss the implications of our constructions on the torus and disk geometry. We also discuss the possibility of generalising our analysis to other topologically ordered states. In Sec.\ref{conclusion}, we make a concluding remark which contains several open problems in the fields and their relations to our work.

\section{The Laughlin state}
\label{Laughlin}
We start with a simple example of the Laughlin state, which is a building block of our construction, to fix the notations and to make our discussion self-contained. The identification between the index of the partition functions and the bulk operators of the Laughlin states can be seen in \cite{Wen:1990se, Milovanovic:1996nj}, and we mainly follow their arguments.

The first quantized $n-$particle wavefunction of the Laughlin ground state at filling factor $\nu =1/q$ with spacial coordinate $\{ z_{i}\}_{i=1}^{n}$ in the plane geometry is given by
\begin{equation}
\Phi (z_{1}, ... ,z_{n})=\left(\prod_{i,j; i<j} \left( z_{i}-z_{j} \right)^{q}\right) e^{-\frac{1}{4}\sum_{i} |z_{i}|^{2}}.
\end{equation}
where we take the magnetic length $\ell_B=\sqrt{\hbar/eB}=1$, with $e$ being the electron charge and $B$ being the magnetic field to avoid complications. In the existing literature, this factor may appear in the exponential part of the wavefunction. With respect to CFTs, one can identify this wavefunction as a bosonic correlation function with a suitable background charge\cite{Moore:1991ks}. It is explicitly given by $\langle \prod_{i} \text{exp}\left(i\sqrt{q} \varphi\left(z_{i}\right)\right) \rangle_{bg}$, where $\varphi$ is the bosonic field and $q$ can be interpreted as the parameter for the $U(1)$ Kac-Moody algebra. It should be noted that this background charge insertion is different from that of the traditional Dotsenko-Fateev Coulomb gas commonly used in the mathematical physics community, because we do not consider the screening charge and the BRST argument\cite{Dotsenko:1984nm,Dotsenko:1984ad}. Hence the resulting theory is still $c=1$ and the fusion rule is not altered by the background charge.

Next, let us introduce the quasihole operator which is the central object for the bulk TO. The quasihole operator at position $w$ for the above wavefunction is $\prod_{i}(z_{i}-w)$. Hence by inserting this operator repeatedly at the same location, we can obtain quasihole wavefunctions characterized by the charge $\frac{1}{q} ,..., 1-\frac{1}{q}, 1$. Insertion of this operator $q$ times corresponds to adding one hole or the removal of one electron. The quasihole states correspond to the primary fields of $U(1)$ Kac-Moody algebra, $e^{ir\phi/\sqrt{q}}$ with $r=0,..., q-1$.

To preserve the conformal symmetry, we demand the quasihole energies to be proportional to their angular momentum (i.e. the eigenvalues of $\sum z_{i}\partial_{z_{i}}$)\cite{Milovanovic:1996nj} By introducing the parameter $s_{N}=\sum_{i} z_{i}^{N}$, the energies of the quasihole states $\prod_{\alpha} s_{N_{\alpha}}\Phi$ are given by $v_F\sum _{\alpha} N_{\alpha}$, where $v_F$ is the Fermi velocity. In each angular momentum sector, the quasihole degeneracies are given by the integer partition function of the angular momentum.

Next, we introduce the excitation which corresponds to adding electrons to the edge. This corresponds to the insertion of $U(1)$ charge for each edge\footnote{It is worth to note that this charged excitation is gapped in cylinder geometry in the lattice models, but to make the discussion simpler we have introduce here.}. Thus the corresponding partition function for the chiral edge is given by:
\begin{equation}
Z=\sum_{r=0}^{q-1}\theta_{\frac{r}{q}}^{+}(x).
\end{equation}
where the chiral character with modular parameter $\tau=iv_F\beta$ is defined by the inverse temperature $\beta$ containing the energy scale $v_{F}$:
\begin{equation}
\theta_{\frac{r}{q}}^{+}=\frac{1}{\eta(x)}\sum_{m=-\infty}^{\infty} x^{\frac{\left(qm+r\right)^{2}}{2q}}, \ x=e^{2\pi i\tau},
\end{equation}
where $r$ characterizes the number of the quasihole and $m$ characterizes the $U(1)$ charge insertion or the particle number, which is interpreted as the fermionic parity in the later sections. The inverse of $\eta$ function represents the contribution from the descendant operators, with $\eta(x)=x^{-\frac{1}{24}}\prod_{n=1}^{\infty}{\left(1-x^{n}\right)}$ derived from the quasihole degeneracies in each angular momentum sector. 

In cylinder geometry with two edges, the analysis can be easily extended by replacing $z_{i}\rightarrow Z_{i}=e^{2\pi  i\frac{z_{i}}{L}}$, and introducing quasihole operator as $\prod_{i} Z_{i}$ with $s_{N}=\sum_{i} Z_{i}^{N}$ and $\overline{s}_{N}=\sum_{i} Z_{i}^{N}$. Starting from the CFT vacuum, and applying the operators corresponding to these operations, we can obtain the Hilbert space as
\begin{equation}
\oplus_{r=0}^{q-1}V_{\frac{r}{q}}\times \overline{V}_{\frac{r}{q}},
\end{equation}
where $V_{\frac{r}{q}}$ and $\overline{V}_{\frac{r}{q}}$ correspond to the chiral and antichiral character $\theta_{\frac{r}{q}}^{+}$, $\overline{\theta}_{\frac{r}{q}}^{+}$ for each of the two edges of the cylinder respectively. Taking into account all of the above discussions, we obtain the cylinder partition function as:
\begin{equation}
Z=\sum_{r=0}^{q-1}\theta_{\frac{r}{q}}^{+}(x) \overline{\theta}^{+}_{\frac{r}{q}}(\overline{x}).
\label{boson}
\end{equation}
As we will discuss in the next section, this partition function satisfies the modular invariance, and this property is significant in considering its relation to the TQFT which should satisfy modular invariance\cite{Cappelli:1996np}.

For later use to specify the $Z_{2}$ fermionic parity, we also introduce the following character,
\begin{equation}
\theta_{\frac{r}{q}}^{-}=\frac{1}{\eta(x)}\sum_{m=-\infty}^{\infty}\left(-1\right)^{m} x^{\frac{\left(qm+r\right)^{2}}{2q}}.
\label{theta_pl}
\end{equation}

Thus the parity even and odd characters are,
\begin{align}
\theta_{\frac{r}{q}}^{\text{even}}&=\frac{1}{2}\left( \theta_{\frac{r}{q}}^{+}+\theta_{\frac{r}{q}}^{-}\right), \\
\theta_{\frac{r}{q}}^{\text{odd}}&=\frac{1}{2}\left( \theta_{\frac{r}{q}}^{+}-\theta_{\frac{r}{q}}^{-}\right). 
\end{align}

\subsection{Modular transformation in $2+1$ dimensional cylinder}

The modular transformation in the FQHE state was first introduced by Cappelli and Zemba\cite{Cappelli:1996np,Cappelli:1998ma,Cappelli:2010jv,Cappelli:2013iga}, and studied by Read and Ino for the FQHE states in a constructive approach\cite{Milovanovic:1996nj,Ino:1998by}. There exist similar analyses in the condensed matter community with applications to more general topological ordered systems, such as the symmetry protected topological ordered system\cite{Ryu_2012,Chen_2017,Chen_2017_2,Chen_2016}.
Let us introduce the general partition function of the FQHE which is constructed by coupling a CFT with the Laughlin state as:
\begin{eqnarray}
Z(\tau ,\zeta)=\text{Tr}[ e^{2\pi i\tau H_{CFT\times U(1)}+2\pi i \zeta Q_{U(1)}}]
\end{eqnarray}
where $H_{CFT\times U(1)}$ is the Hamiltonian of the corresponding to the CFT and the Laughlin state part; $Q_{U(1)}=m+\overline{m}+r/q+\overline{r}/q$ is the total $U(1)$ charge of the model, and $\zeta$ is the chemical potential corresponding to $Q_{U(1)}$. As we will see in the subsequent sections, the definition of trace in the partition function $Z(\tau ,\zeta)$ differs depending on the literature. However, by introducing the technique of fermionization, we show that the gap between the literature can be fullfilled. The action of the modular transformations are 
\begin{eqnarray}
&&S:Z(\tau ,\zeta)\to Z\left(-\frac{1}{\tau},-\frac{\zeta}{\tau}\right)\\
&&T^{2}:Z(\tau ,\zeta)\to Z(\tau+2 ,\zeta)\qquad \\
&&U:Z(\tau ,\zeta)\to Z(\tau ,\zeta+1)\\
&&V:Z(\tau ,\zeta)\to Z(\tau ,\zeta+\tau)
\end{eqnarray}

Here, we summarize the interpretations of these transformations as follows. The $S$ transformation is the usual low-high temperature dual transformation. The invariance under $T^{2}$ transformation indicates that the system is constructed from either the bosonic or fermionic physical excitations (i.e. either with with integer or half-integer spin). On the other hand, the $U$ transformation indicates the total charge is an integer, and the $V$ transformation represents the addition of a flux quantum. 

The partition function in Eq.(\ref{boson}) correspond to the case where $H_{CFT\times U(1)}$ only has the $U(1)$ part.  Fortunately, the modular transformation property of the $U(1)$ part has already been calculated in \cite{Ino:1998by}.
Here we note the action of modular $S$ transformation to the $U(1)$ part which are relevant for the later discussion:
\begin{equation}
\sum_{r=0}^{q-1}\theta_{\frac{r}{q}}^{+}\left(-\frac{1}{\tau}\right)\overline{\theta}_{\frac{r}{q}}^{+}\left(-\frac{1}{\overline{\tau}}\right)=\sum_{r=0}^{q-1}\theta_{\frac{r}{q}}^{+}(\tau)\overline{\theta}_{\frac{r}{q}}^{+}(\overline{\tau}),
\label{modular_S_Laughlin}
\end{equation}
\begin{equation}
\sum_{r=0}^{q-1}\theta_{\frac{r+1/2}{q}}^{+}\left(-\frac{1}{\tau}\right)\overline{\theta}_{\frac{r+1/2}{q}}^{+}\left(-\frac{1}{\overline{\tau}}\right)
=\sum_{r=0}^{q-1}\theta_{\frac{r}{q}}^{-}(\tau)\overline{\theta}_{\frac{r}{q}}^{-}(\overline{\tau}).
\label{modular_S_Laughlin_2}
\end{equation}
The LHS of Eq.(\ref{modular_S_Laughlin_2}) represents the insertion of a ``half-flux quantum" by modifying the index $r$ to $r+1/2$ in Eq. (\ref{theta_pl}). Hence, as we will see in the next section, one can observe that the half-flux part which appears as characteristic structure of the coupling condition of FQHE breaks the modular $S$ invariance. In our construction based on the fermionic CFT, we can add the effect of $\zeta$ after constructing the modular $S$ invariant partition function. Hence we take $\zeta=0$ for convenience in the following sections. 

\section{Cylinder partition function constructed from fermionic conformal field theory}
\label{cylinder_fermion}

We now proceed to study a general case with a fermionic CFT (which we denote as $FM$) having a non-anomalous $Z_{2}$ symmetry generated by a simple current $J$\cite{Hsieh:2020uwb, Fukusumi:2021zme,Runkel:2020zgg}, and propose the construction of the partition function corresponding to the $CFT_{D}/BQFT_{D+1}$. We assume that the corresponding bosonic theory $M$ can be represented by $Z_{2}$ noninvariant sector labeled by the index $i\pm$ and $Z_{2}$ invariant sector labeled by the index $a$. The signs of $i\pm$ is labeling the fermionic parity, where``$+$" is parity even and ``$-$" is parity odd. For simplicity, we assume that the fermionic theory $FM$ is constructed from bosonic theory $M$ with the bosonic $Z_{2}$ charge conjugated partition functions\footnote{However, one can apply our construction to general models, by considering the fermionic parity, simple charge and fermionic zero modes as we discuss in the follwoing sections.}:
\begin{equation}
Z_{M}=\sum_{i} \left(|\chi_{i+}|^{2}+|\chi_{i-}|^{2}\right)+\sum_{a}|\chi_{a}|^{2},
\end{equation}
where $\chi_{i\pm}=\chi_{i\pm}(\tau)$ and $\chi_{a}=\chi_{a}(\tau)$ are the chiral character labeled by $i\pm$ and $a$ respectively.

Let us introduce the construction of the electron operator from the $Z_{2}$ simple current $J$, which is a building block of the FQHE wavefunctions. The $Z_{2}$ simple current satisfies the fusion rule $J\times \phi_{\alpha}=\phi_{\alpha'}$, where $\alpha$ and $\alpha'$ are labels of the fields, from which we use the notation $\alpha'=J(\alpha)$. The earlier attempts focused on the simple current of CFT\cite{Frohlich:2000qs,Cappelli:2010jv}, and one can represent the electron operator as $\Psi=Je^{i\sqrt{q}\varphi}$, where $\varphi$ corresponds to the $U(1)$ part as introduced in the previous sections. Hence to consider the CFT operator corresponding to the quasihole excitation, one can consider the operator product expansion (OPE) with the  
$\Psi$. The general OPE with the field $\phi_{\alpha}$ (where $\alpha$ is a label of the fields and takes $i\pm$ or $a$ respectively), with the current $J$ is:
\begin{equation}
\phi_{\alpha}(z)\times J(z')\sim \frac{1}{(z-z')^{h_{\alpha}+h_{J}-h_{J(\alpha)}}}\phi_{J(\alpha)} (z'), 
\end{equation}
where $h_{\alpha}, h_{J}, h_{J(\alpha)}$ are the conformal dimension of the fields labeled by $\alpha, J, J(\alpha)$ respectively.

It is natural to introduce the following simple charge for the operator labeled by $\alpha$,
\begin{equation}
Q_{J}(\alpha)=h_{\alpha}+h_{J}-h_{J(\alpha)},
\end{equation}
which is an integer or a half-integer. We denote the half-integer case as ``twisted" and integer case as ``untwisted", which corresponds to the NS and R sector respectively\footnote{For the readers interested in the structure of the correlation functions in fermionic CFT, see\cite{Petkova:1988cy,Furlan:1989ra}.}. The single-valuedness of the wavefunctions results in a different set of $U(1)$ parts corresponding to this simple charge, and it results in a half-flux insertion $r\rightarrow r+1/2$ for the twisted characters. In this section, we apply the notation where $r$ only takes integer values, and by coupling the fermionic CFT to $U(1)$, we get the character functions $\Xi$ with the total fermionic parity as:
\begin{align}  
\Xi_{i \pm,\frac{r}{q}}^{\text{even}}&= \frac{1}{2}\chi_{i\pm}\left( \theta_{\frac{r}{q}}^{+}\pm\theta_{\frac{r}{q}}^{-} \right),\\
\Xi_{i \pm,\frac{r}{q}}^{\text{odd}}&= \frac{1}{2}\chi_{i\pm}\left( \theta_{\frac{r}{q}}^{+}\mp\theta_{\frac{r}{q}}^{-} \right),\\
\Xi_{i \pm,\frac{r}{q}}^{\text{tw, even}}&= \frac{1}{2}\chi_{i\pm}\left( \theta_{\frac{r+1/2}{q}}^{+}\pm\theta_{\frac{r+1/2}{q}}^{-} \right),\\
\Xi_{i \pm,\frac{r}{q}}^{\text{tw, odd}}&= \frac{1}{2}\chi_{i\pm}\left( \theta_{\frac{r+1/2}{q}}^{+}\mp\theta_{\frac{r+1/2}{q}}^{-} \right),
\end{align}
where $\Xi_{a, \frac{r}{q}}=\chi_{a}\theta^{+}_{\frac{r+1/2}{q}}$ for $\ h_{J}=\text{half-integer}$, $\Xi_{a, \frac{r}{q}}=\chi_{a}\theta^{+}_{\frac{r}{q}}$ for $\ h_{J}=\text{integer}$ for the $Z_{2}$ invariant sectors.
It should be noted that the index $a$ does not contain the fermionic parity. This is the consequence of $J\phi_{a}=\phi_{a}$, where $\phi_{a}$ is the $Z_{2}$ invariant sector. In other words, this operator itself should be treated as the zero modes under the fermionization. These zero modes correspond to the order and disorder operators in the statistical mechanical models\cite{Ginsparg:1988ui}, and one can thus represent this operator as $\phi_{a}=\frac{\phi_{a}^{\text{even}}+\phi_{a}^{\text{odd}}}{\sqrt{2}}$. More precisely, there exist gauge choice, $\phi_{a}=\frac{\pm\phi_{a}^{\text{even}}\pm\phi_{a}^{\text{odd}}}{\sqrt{2}}$, but we have chosen the simplest gauge which results in the positive integer matrices representations\cite{Chang:2022hud}. The other choice $\mu_{a}=\frac{\phi_{a}^{\text{even}}-\phi_{a}^{\text{odd}}}{\sqrt{2}}$ can appear as a disorder operator, and one can distinguish $\phi_{a}$ and $\mu_{a}$ by implementing fermion parity operator $(-1)^{F}$. \footnote{We thank Yuji Tachikawa for the useful comments for this problem and Yunqin Zheng for sharing the manuscipt \cite{Li:2023mmw} containing related topics.}. Generally, one can relate a bosonic topological order to the modular tensor category (MTC), but its fermionized counter part is still under development\cite{Lan_2016,Aasen:2017ubm,Inamura:2022lun}. Moreover, there may exist some difficulties for the researchers in other fields to follow their approaches because of the abstract aspects of category theories and their mathematical complications. The representation $\frac{\phi_{a}^{\text{even}}\pm\phi_{a}^{\text{odd}}}{\sqrt{2}}$ gives an intuitive understanding of the fermionized representation of fusion rule in the fermionized MTC, based on the $Z_{2}$ simple charge and fermionic parity.

By considering all pattern of the excitations as in the Laughlin case, and introducing the fermionic parity of the total systems, a cylinder partition function with even parity without bulk zero modes is:
\begin{equation}
\begin{split}
Z_{T^{2}-\text{inv}}^{\text{even}}&=\sum_{i, r}\left(|\Xi_{i+,\frac{r}{q}}^{\text{even}}+\Xi_{i-,\frac{r}{q}}^{\text{even}}|^{2}+|\Xi_{i+,\frac{r}{q}}^{\text{odd}}+\chi_{i-,\frac{r}{q}}^{\text{odd}}|^{2} \right) \\
&+\sum_{a, r}|\Xi_{a,\frac{r}{q}}|^{2}
\end{split}
\end{equation} 
which is invariant under the modular $T^{2}$ transformation; $i$ and $a$ take all value, and we have also assumed $i+$ and $i-$ belong to the same twisted or untwisted sector because the theory is nonanomalous. The corresponding odd-parity partition function $Z_{T^{2}-\text{inv}}^{\text{odd}}$ can be obtained by exchanging the odd and even part of the holomorphic part of the even-parity partition function, summarized as the transformation $\Xi^{\text{odd}}\leftrightarrow \Xi^{\text{even}}$. This exchange of the odd and even parity can be thought of as a kind of parity shift operation in \cite{Runkel:2020zgg}. The form of $T^{2}$ invariant partition can be changed depending on whether the conformal dimension of the simple current operator is half-integer or integer. 

Hence the total partition function which corresponds to $CFT_{D}/BQFT_{D+1}$ correspondence is,
\begin{equation}
\begin{split}
Z_{T^{2}-\text{inv}}&=Z_{T^{2}-\text{inv}}^{\text{even}}+Z_{T^{2}-\text{inv}}^{\text{odd}} \\
&=\sum_{i, r}\left(|\Xi_{i+,\frac{r}{q}}+\Xi_{i-,\frac{r}{q}}|^{2}\right)
+2\sum_{a, r}|\Xi_{a,\frac{r}{q}}|^{2}.
\end{split}
\label{T2inv}
\end{equation}
where we have introduced the notation $\Xi_{i\pm}=\Xi_{i\pm}^{\text{even}}+\Xi_{i\pm}^{\text{odd}}$ as for $\Xi_{a}$. Remarkably, one can apply the following decomposition to the partition function $Z_{T^{2}-\text{invariant}}$,
\begin{equation}
Z_{T^{2}-\text{inv}}=Z_{T^{2}-\text{inv}}^{\text{untw}}+Z_{T^{2}-\text{inv}}^{\text{tw}}
\end{equation}
where $Z_{T^{2}-\text{inv}}^{\text{untw}}$ and $Z_{T^{2}-\text{inv}}^{\text{tw}}$ represent the ``untwisted" (i.el NS) and ``twisted" (i.e. R) sector respectively, determined by the simple charge $Q_{J}(\alpha)$ of each sector $\alpha$ \footnote{As one of the author discusses in the related works\cite{Fukusumi_2022, Fukusumi_2022_c}, this decomposition is a remarkable property of nonanomalous conformal dimension of the $Z_{2}$ simple current}. One can observe the modular $S$ non-invariance of the partition function in Eq.\eqref{T2inv}, by applying modular $S$ invariance of the NS sector and modular $S$ non-invariance of the R sector. Because the partition function in Eq. \eqref{T2inv} is lacking modular $S$ invariance, we have interpreted this bulk-edge correspondence as $CFT_{D}/BQFT_{D+1}$ correspondence, not as $CFT_{D}/BTQFT_{D+1}$ correspondence. 

\subsection{The Modular invariant part}
Here we introduce an argument for the modular property of the total partition function. The procedure focuses on a particular part of the total partition function with the introduction of the imaginary gapping operation. Typically, we focus on the modular $S$ property of NS and R sector attached with flux. We do not require this imaginary gapping operation to protect the gaplessness of the edge modes. It should be stressed that our formalism, regardless of its simplicity, can reveal the properties of the excitations systematically and visibly for the interpretation of the bulk and boundary RG as we show in the following discussion.

For $Z_{T^{2}-\text{inv}}$, we consider gapping out the half-flux quantum sector which breaks modular $S$ invariance. This modular non-invariance can be observed by the orbifolding construction of the R sector and the modular $S$ non-invariance of the Laughlin part of half-flux quantum sector Eq. \eqref{modular_S_Laughlin_2}.  Then, we can obtain the fermionic partition function which is equavalent to $Z_{T^{2}-\text{inv}}^{\text{untw}}$ as: 
\begin{equation}
Z_{T^{2}-\text{inv}}\rightarrow\sum_{i, r\in \text{untwisted}}\left(|\Xi_{i+,\frac{r}{q}}+\Xi_{i-,\frac{r}{q}}|^{2}\right)
\end{equation}
for $h_{J}=\text{half-integer}$ and,
\begin{equation}
Z_{T^{2}-\text{inv}}\rightarrow\sum_{i, r\in \text{untwisted}}\left(|\Xi_{i+,\frac{r}{q}}+\Xi_{i-,\frac{r}{q}}|^{2}\right)+2\sum_{a,r}|\Xi_{a,r}|^{2}
\end{equation}
for $h_{J}=\text{integer}$. This is nothing but the product of the partition function of the NS sector of the fermionic CFT and that of the Laughlin states\cite{Hsieh:2020uwb, Fukusumi:2021zme}. It is consistent with the result in \cite{Cappelli:2010jv}, and our analysis filled the gap between the result of \cite{Milovanovic:1996nj,Ino:1998by} and \cite{Cappelli:2010jv}. Hence we can conclude these fermionic excitations satisfy a necessary condition for stability, which is the modular $S$ invariance from the orbifolding construction of the fermionic theories\cite{Hsieh:2020uwb}. This appearance of $Z^{\text{untw}}_{T^{2}-\text{inv}}$ may correspond to the spontaneous supersymmetry breaking in \cite{Ma:2021dua,Nguyen:2022khy} which captured the attentions in the high energy physics community as a possible basic phenomenon in particle physics. In short, we have proposed a procedure obtaining the partition function corresponding to $CFT_{D}/BTQFT$ starting from $CFT_{D}/BQFT_{D+1}$ correspondence, known as the bulk-edge correspondence.

Moreover, these partition functions correspond to the $\left(2+1\right)-$dimensional topologically ordered states also in the bulk by considering the mapping to the tunnel proposed in \cite{Wen:1990se} as we will discuss in the next section. It should be noted that this fermionic (or electronic) construction, which is a central subject in condensed matter physics, can lead to the appearance of a series of possibly non-abelian particles. In our construction, the structure of the Lagrangian subalgebra appears naturally as the untwisted part of the fermionic theory, because the untwisted or simple charge zero fields $\{ \phi_{\alpha}^{\text{untw}}\}$ produce the following fusion rules:

\begin{equation}
\phi_{\alpha_{1}}^{\text{untw}}\times \phi_{\alpha_{2}}^{\text{untw}}=\sum_{\gamma} N_{\alpha_{1},\alpha_{2}}^{\gamma} \phi_{\gamma}^{\text{untw}},
\end{equation}
where $N_{\alpha_{1},\alpha_{2}}^{\gamma}$ is the fusion matirx of the theory.

Hence on the cylinder, this construction gives a systematic way of obtaining the protected edge modes.
In the forth coming paper\cite{Fukusumi_2022_c}, we will show this structure is universal for the general topological ordered systems. By considering the simple charge of fields in CFTs and their untwisted parts, a class of Lagrangian subalgebra can be obtained. By applying the imaginary gapping operation only to the bosonic or Laughlin state parts, we can also obtain the general fermionic CFT partition functions. There exist similar arguments in \cite{Cheng:2022nji,Cheng:2022nds}. 

It is important to note that our analysis shows a difficulty in obtaining a bosonic partition function $Z_{M}$ and the corresponding bosonic TO started from the non-Abelian FQHE (FIG. \ref{boson_fermion_fqhe}) (for the readers interested in the details, see the discussion in Appendix \ref{bosonic_pf}). This connectivity analysis to the FQHE can contrain the RG connectivity of the phase diagram. Moreover, this can be thought of as a generalized (and more accurate) argument of RG connectivity based on the modular property in \cite{Furuya:2015coa} by combining the $Z_{2}$ (more generally $Z_{N}$) anomaly classification in \cite{Tachikawa:2017gyf,Bhardwaj:2017xup,Fukusumi_2022,Fukusumi_2022_c}.

\begin{figure}[htbp]
\begin{center}
\includegraphics[width=0.5\textwidth]{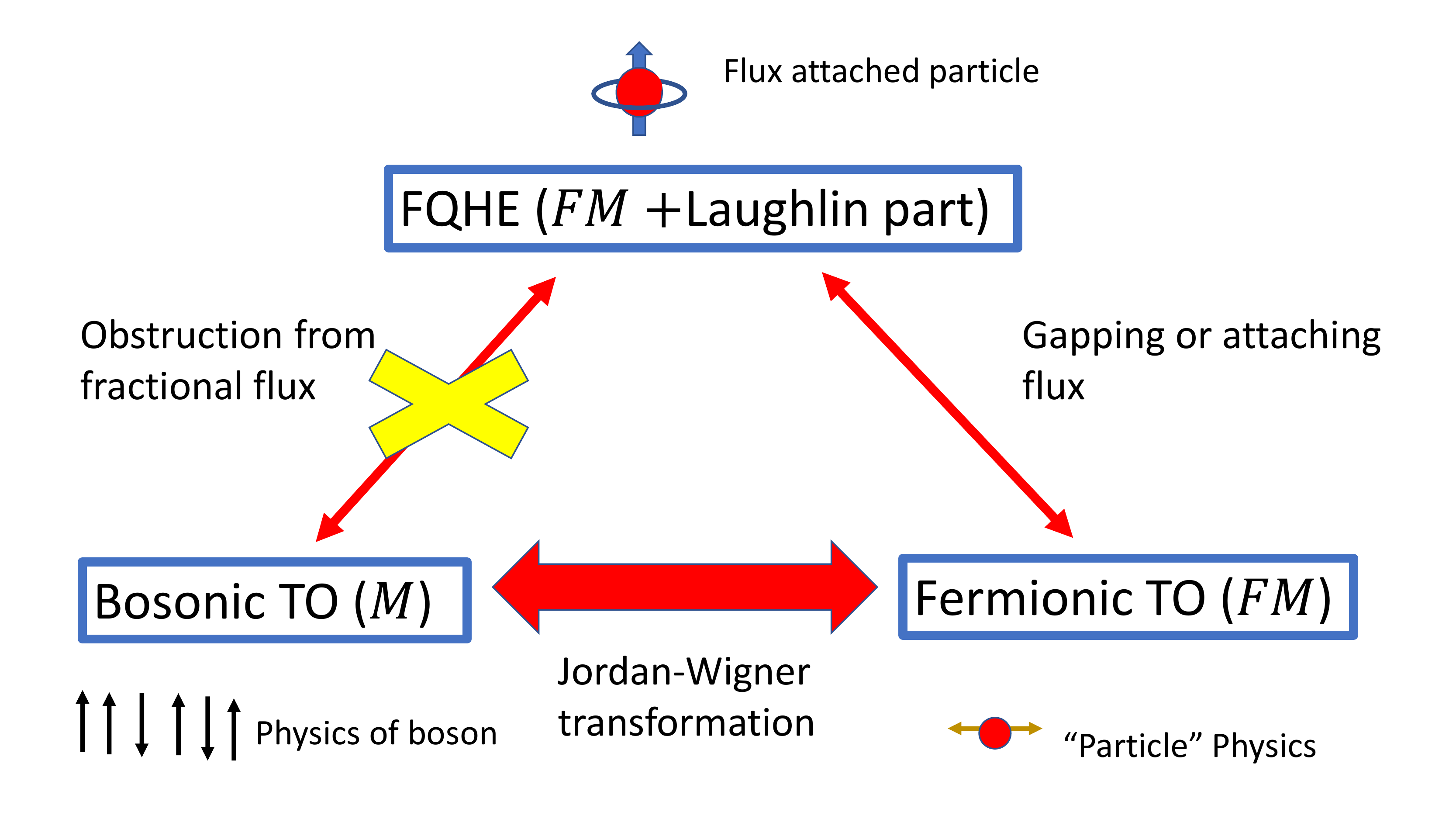}
\caption{Connectivity diagram of bosonic and fermionic TO to FQHE. The bosonic TO ($M$) and fermionic TO ($FM$) are related by the Jordan-Wigner transformation. It should be worth noting that this transformation acts nonlocally on the Hilbert space. By this transformation, the bosonic system, such as a spin system, is transformed into the fermionic particle system. By attaching $U(1)$ flux corresponding to this particle system, one can obtain FQHE. In this sense, FQHE should be interpreted as a kind of gauge theory.}
\label{boson_fermion_fqhe}
\end{center}
\end{figure}

Before going to the next section, we give a brief comment on the reason why the Laughlin state is so ubiquitous even when considering gapless FQHE such as the Gaffnian CFT\cite{Weerasinghe_2014,Regnault2009TopologicalEA,Estienne2015CorrelationLA,Yuzhu_2023}. It is easy to obtain a modular invariant, Eq. (\ref{boson}), by gapping the CFT and the fractional flux parts except for the Luttinger liquid. In this sense, one can expect that the Laughlin state is more robust or universal.

\subsection{Modular invariance of Moore-Read Pfaffian state and its fermionic duality}
\label{Moore_Read}

In this subsection, we review the Moore-Read Pfaffian states from a modern perspective. Usually, we assume the following fusion rule:
\begin{align}
\epsilon\times \epsilon &= I, \\
\epsilon \times \sigma &=\sigma, \\
\sigma \times \sigma &=I+\epsilon,
\end{align}
where $I$ is the identity operator, $\epsilon$ is the $Z_{2}$ simple current, and $\sigma$ is the spin operator respectively. Their respective conformal dimensions are $h_{I}=0$ $h_{\epsilon}=1/2$ and $h_{\sigma}=1/16$.
As we have discussed in the previous sections, there exist zero modes for $\sigma$ because of its $Z_{2}$ invariance. This is a consequence of the obstruction of the implementation of the fermionic parity in the bosonic theories. If we set the fermionic parity of the identity operator as even and that of $\epsilon$ as odd, we have to introduce the zero modes for $\sigma=\frac{\sigma^{\text{even}}+\sigma^{\text{odd}}}{\sqrt{2}}$ to establish the compatibility of the fusion rule and the fermionic parity of the system. For the reason which becomes clear later, we introduce the notation $\sigma^{\text{even}}=e$, and $\sigma^{\text{odd}}=m$. The new fusion rule on this basis is,

\begin{align}
\epsilon\times \epsilon &= I, \\
\epsilon \times m &=e, \\
\epsilon \times e &=m, \\
e \times e &=m \times m =I, \\
m \times e &=\epsilon.
\end{align}
This is nothing but the fusion rule for the semions and the Majorana fermions\cite{Kitaev:2006lla}. One can generalize this correspondence to the $Z_{N}$ simple currents. Mathematically, this may give another representation of the Tambara-Yamagami category.  

The Pfaffian states are described by the simplest $T^{2}$ invariant form in the previous section:
\begin{equation}
\begin{split}
Z_{T^{2}-\text{inv}}^{\text{even}}&=\sum_{r}\left(|\Xi_{0,\frac{r}{q}}^{\text{even}}+\Xi_{\epsilon,\frac{r}{q}}^{\text{even}}|^{2}+|\Xi_{0,\frac{r}{q}}^{\text{odd}}+\Xi_{\epsilon,\frac{r}{q}}^{\text{odd}}|^{2} \right), \\
&+\sum_{r}|\Xi_{\sigma,\frac{r}{q}}|^{2}
\end{split}
\end{equation}
\begin{equation}
\begin{split}
Z_{T^{2}-\text{inv}}^{\text{odd}}&=\sum_{r} \left( \Xi_{0,\frac{r}{q}}^{\text{odd}}+\Xi_{\epsilon,\frac{r}{q}}^{\text{odd}}\right)\left(\overline{\Xi}_{0,\frac{r}{q}}^{\text{even}}+\overline{\Xi}_{\epsilon,\frac{r}{q}}^{\text{even}}\right) \\
&+\sum_{r}\left( \Xi_{0,\frac{r}{q}}^{\text{even}}+\Xi_{\epsilon,\frac{r}{q}}^{\text{even}}\right)\left(\overline{\Xi}_{0,\frac{r}{q}}^{\text{odd}}+\overline{\Xi}_{\epsilon,\frac{r}{q}}^{\text{odd}}\right) \\
&+\sum_{r}|\Xi_{\sigma,\frac{r}{q}}|^{2},
\end{split}
\end{equation}
\begin{equation}
\begin{split}
Z_{T^{2}-\text{inv}}&=Z^{\text{even}}+Z^{\text{odd}} \\
&=\sum_{r}\left(|\Xi_{0,\frac{r}{q}}+\Xi_{\epsilon,\frac{r}{q}}|^{2}\right)+2\sum_{r}|\Xi_{\sigma,\frac{r}{q}}|^{2}.
\end{split}
\end{equation}
It should be noted that the total partition function is not modular invariant whereas the even part is\cite{Milovanovic:1996nj,Ino:1998by}. However, by gapping the half-flux part, one can obtain the partition function $\sum_{r}\left(|\Xi_{0,\frac{r}{q}}+\Xi_{\epsilon,\frac{r}{q}}|^{2}\right)$. This is nothing but the product of the partition functions of Luttinger liquid and Kitaev chain which is modular $S$ invariant. Hence, in this case, there exist at least two potential procedures to obtain the modular invariance corresponding to the topologically protected edges. One approach is to gap out the parity odd sectors and the other is to gap out the half-flux sectors\footnote{To distinguish between these two cases, it may be reasonable to consider the generalized Gibs partition function and their response theories.}. 

\begin{figure}[htbp]
\begin{center}
\includegraphics[width=0.5\textwidth]{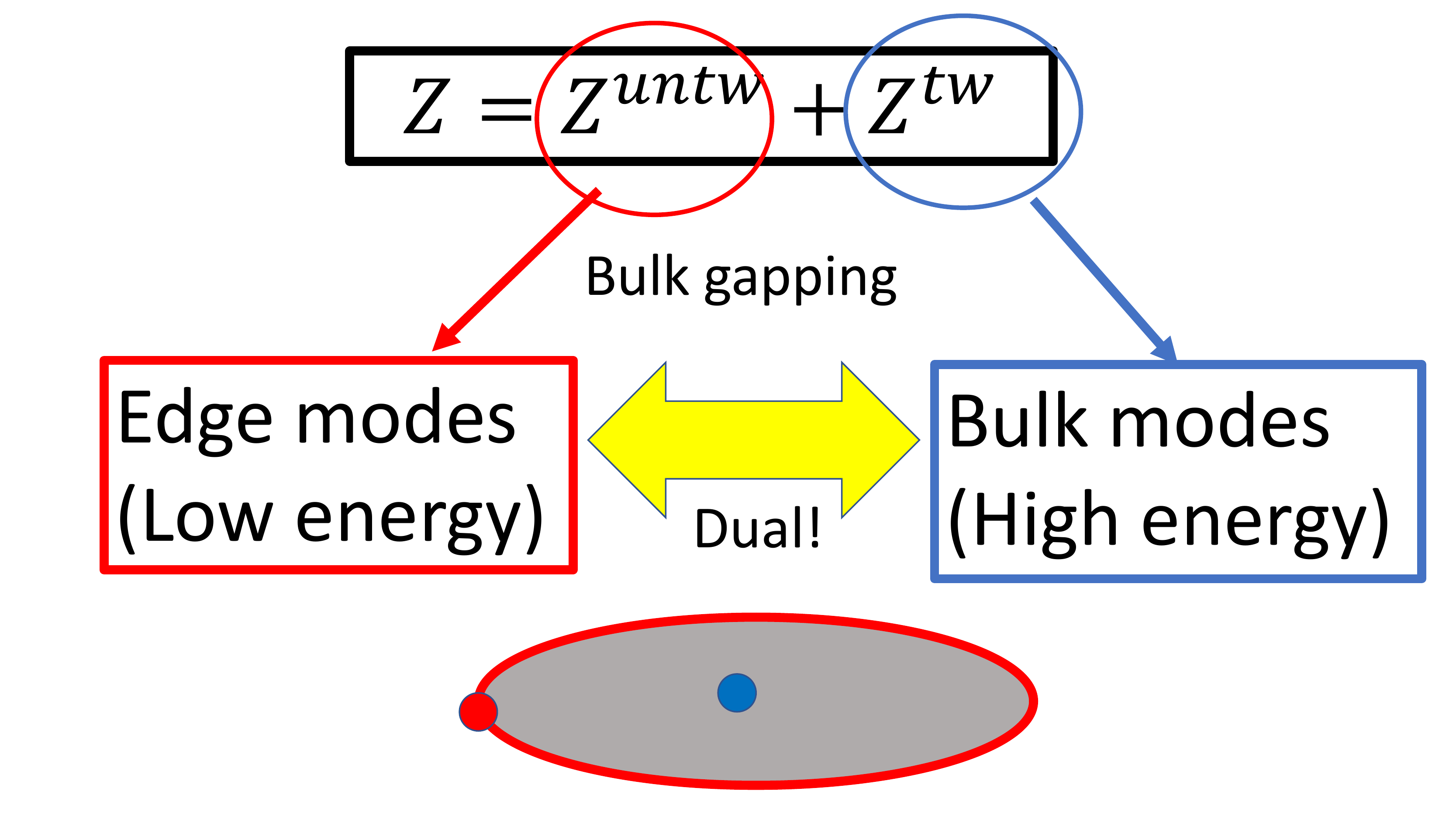}
\caption{Duality interpretation of the partition functions. One can see the mapping of Hilbert space by adding the quasiparticle corresponding to the duality of the bosonic CFT.}\label{duality_fig}
\end{center}
\end{figure}

Before going to the next section, we comment on the duality of the model. Recently, the electromagnetic duality of the Hofstadter model has captured attentions among the condensed matter \cite{Fendley_1995,Fendley:1998sq,Lesage:1998jky} and high energy physicists\cite{Hatsuda:2016mdw,Ikeda:2017ztr}\footnote{This duality is related to the (hidden) quantum group symmetry of CFT and the corresponding lattice model\cite{Gomez:1992bj}. More mathematically, it may correspond to the Langlands program\cite{Frenkel:2005pa,Ikeda:2018tlz}.}. When there exists a $Z_{2}$ duality defect in our fermionic modular $S$ noninvariant partition function Eq. \eqref{T2inv}, one can see a fermionic analog of such duality, which relates the twisted state and untwisted states. In the Moore-Read states, by applying $(\sigma \otimes\overline{\sigma})/2$ for the edges (i.e. adding the quasiparticle $\sigma/\sqrt{2}$ for each edge), one can see the following transformation laws by using the fusion rule:
\begin{align}
|\chi_{0}+\chi_{\epsilon}|^{2} &\rightarrow2|\chi_{\sigma}|^{2} , \\
2|\chi_{\sigma}|^{2} &\rightarrow|\chi_{0}+\chi_{\epsilon}|^{2} , 
\end{align}
where we have only noted the Ising CFT part of the theory, but by attaching the half-flux quantum to $\sigma$ or replacing the character $\chi$ to $\Xi$), one can see the same transformation for the Moore-Read states. Hence by interpreting the untwisted sector as the low energy modes and the twisted sector as the high energy modes in the bulk gapping process, this duality can be interpreted as the high and low energy (not temperature) duality (FIG. \ref{duality_fig}). This is analogous to the duality in the dual resonance model which has been proposed long ago, and one can see the same duality for the general fermionic models with $Z_{2}$ duality defect\cite{Frohlich:2006ch,Schweigert:2007wd}, and it can be referred to as fermionic T duality in the recent terminology. There exists a lot of literature about these dual models in high energy physics\cite{Veneziano:1968yb,Mandelstam:1974fq}, and there exist related discussions in the recent study of supersymmetry in FQHE with emphasis on boundary \cite{Bae:2021lvk}, and operator formalism \cite{Sagi:2016slk}. For the readers who are interested in the historical aspects of the duality, we introduce the legacy of Olive by Corrigan and Goddard\cite{https://doi.org/10.48550/arxiv.2009.05849}, because it contains historical aspects of quark models which have a connection with our approach.  

\section{Ishibashi states from Cardy states}
\label{Ishibashi}

\begin{figure}[htbp]
\begin{center}
\includegraphics[width=0.5\textwidth]{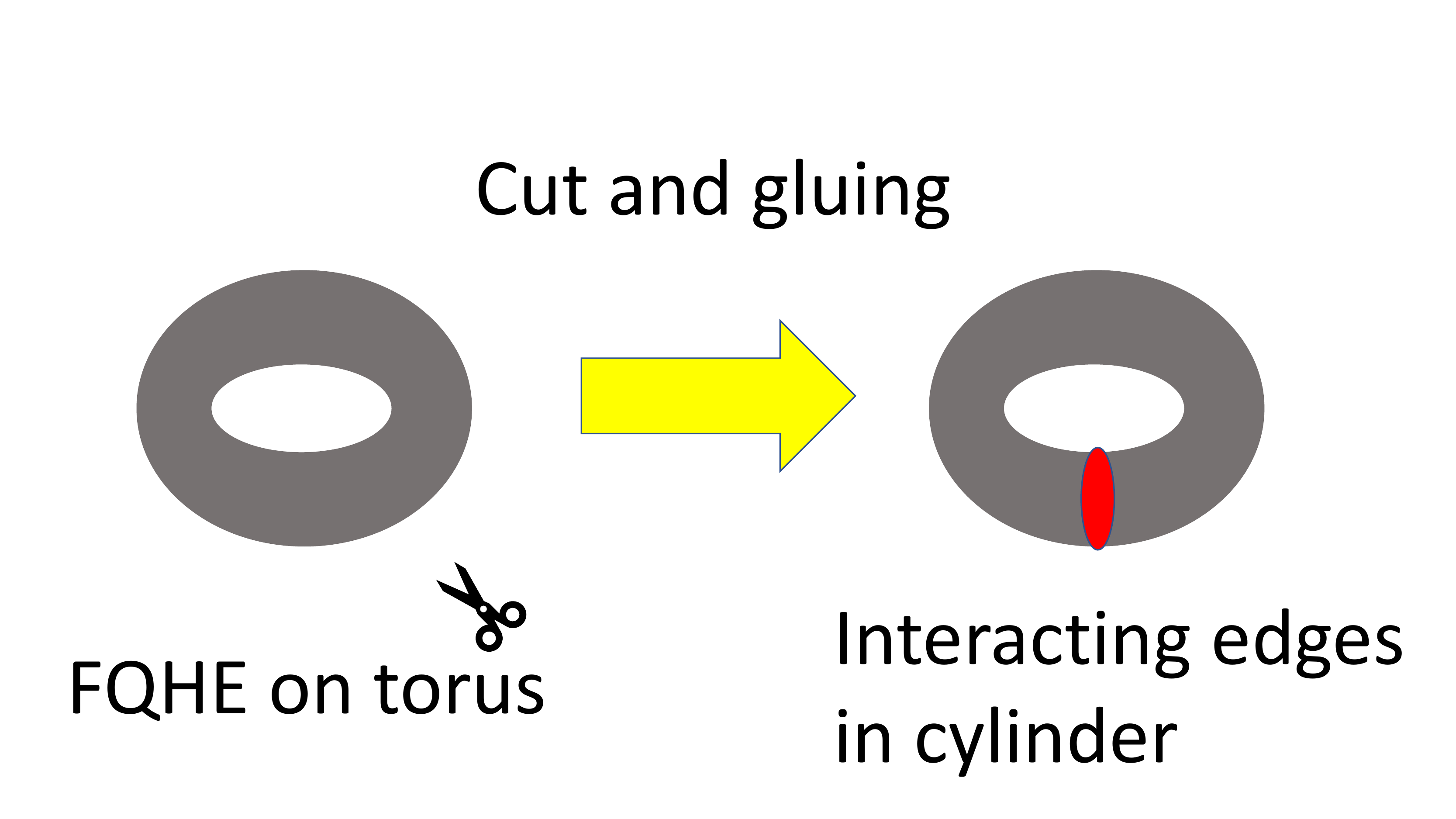}
\caption{Mapping of tunnel problem. Based on the cut and gluing operation, one can interpret the ground state of FQHE as the ground states of interacting edges, corresponding to the full CFT with relevant perturbations.}
\label{tunnel}
\end{center}
\end{figure}

\begin{figure}[htbp]
\begin{center}
\includegraphics[width=0.5\textwidth]{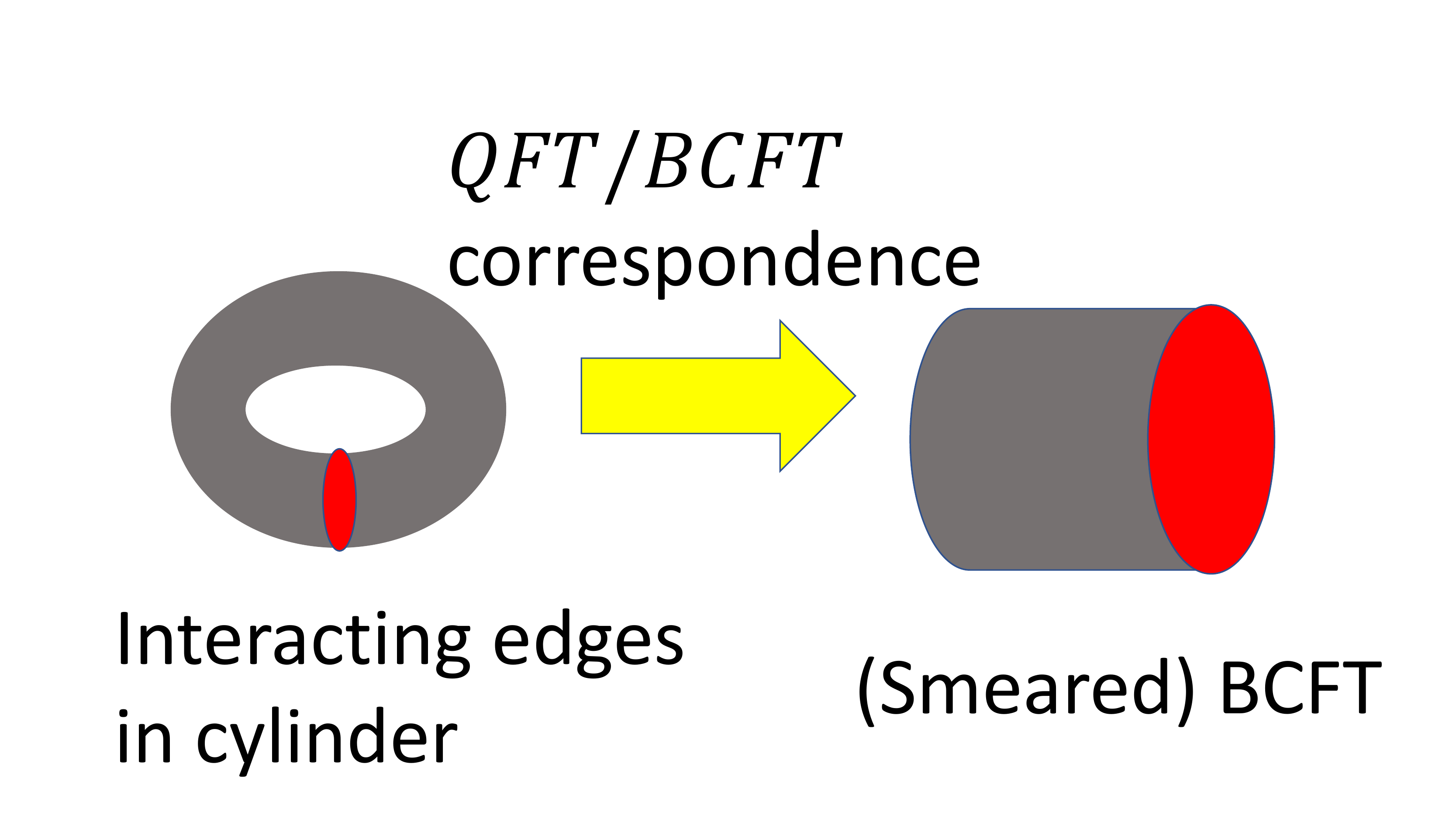}
\caption{QFT/BCFT correspondence. By using the correspondence proposed in \cite{Cardy:2015xaa}, one can map the low energy physics of interacting edges to the physics of (smeared) BCFT, typically Cardy states.}
\label{folding}
\end{center}
\end{figure}

Assuming the bulk-edge correspondence, one expects to map topologically ordered states on a torus to the tunnel problem on a cylinder\cite{Wen:1990se}(FIG. \ref{tunnel}). Related to this, a nontrivial conjecture by Cardy which relates the BCFT to the RG flow of CFT has appeared \cite{Cardy:2017ufe,Lencses:2018paa}. One can name this correspondence as the (massive) QFT/BCFT correspondence in the same space-time dimensions. It was proposed that CFT with relevant perturbation, which is equivalent to the tunnel problem in FQHE, can be analyzed by the smeared boundary states (FIG. \ref{folding})\footnote{There exist some historical aspects which have not captured enough attention in condensed matter physics community related to the phase diagram of integrable models. One can see this correspondence between massive QFT and BCFT in \cite{Date:1987zz,Saleur:1988zx}, for example. For a review of this aspect, see the discussion in \cite{Foda:2017vog}. As is emphasized in \cite{Cardy:2017ufe}, such smeared boundary condition can appear in the quantum quench problem\cite{Calabrese:2006rx,Cardy:2015xaa,Guo:2017rji}.}.

\begin{figure}[htbp]
\begin{center}
\includegraphics[width=0.5\textwidth]{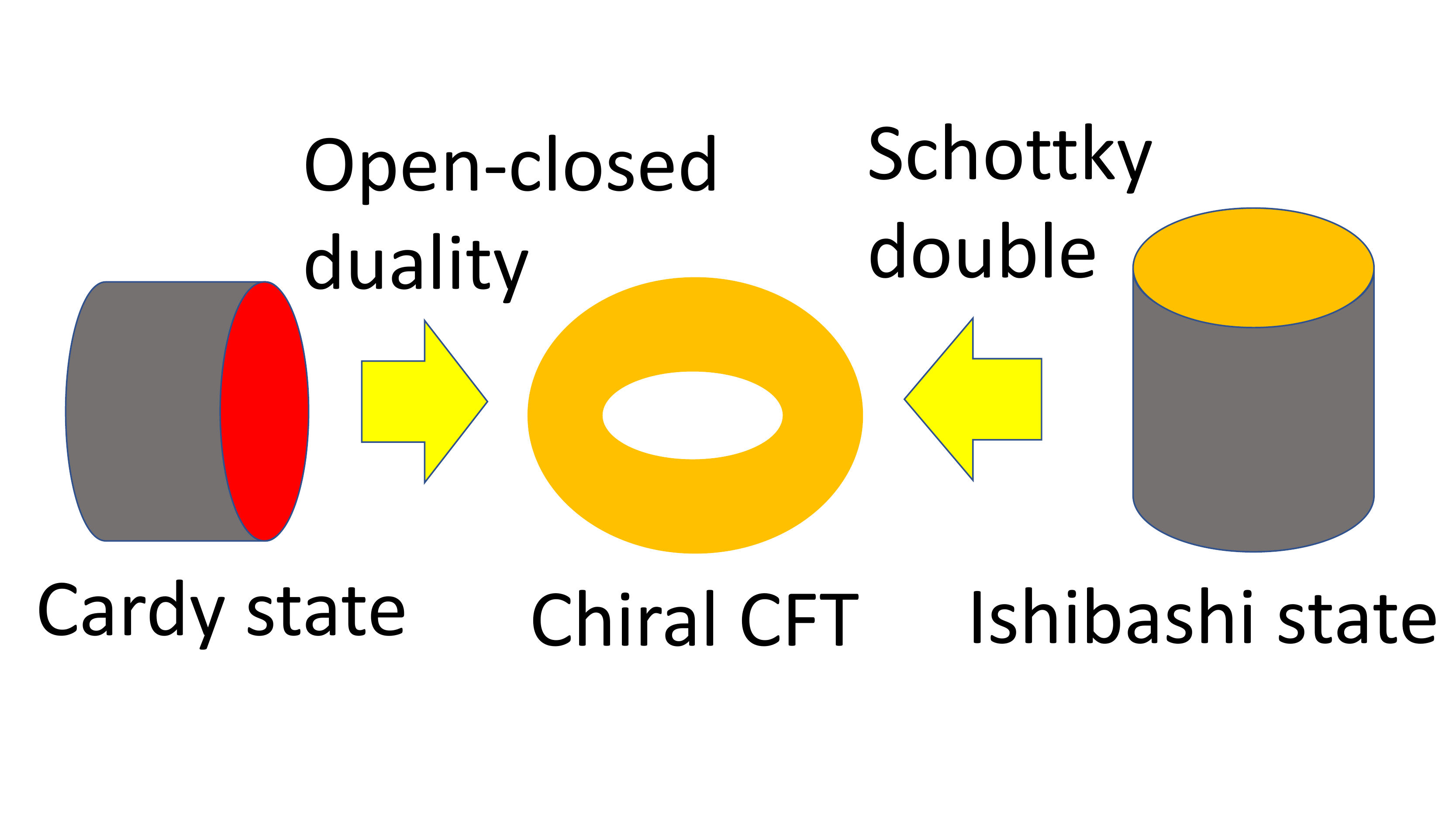}
\caption{Open-closed duality and Schottky double. By using open-closed duality, the (low temperature) Cardy states corresponding to quantum states are mapped to the (high temperature) density matrix of chiral CFT, which can be mapped to Ishibashi states. This correspondence between low-temperature physics and high-temperature physics is one of the most important aspects in studying the entanglment of the systems.}
\label{Schottky}
\end{center}
\end{figure}

Based on this modern understanding, it may be useful to express the cylinder partition function of the Cardy states $\{ |B_{\alpha}\rangle_{\text{Cardy}}\}$ as the amplitude of the Ishibashi states $\{ |\alpha\rangle_{\text{Ishibashi}}\}$ where $\alpha$ is the primary fields of the theory \cite{Das:2015oha,Qi_2012, Nakayama:2015mva}:

\begin{equation}
\begin{split}
\langle B_{\alpha}|e^{2\pi i \tau H_{\text{CFT}}}|B_{\alpha}\rangle_{\text{Cardy}}&=\sum_{\gamma \in \alpha \times \alpha} \chi_{\gamma} (-\frac{1}{\tau}) \\
&=\sum_{\gamma \in \alpha\times \alpha}\langle \gamma|e^{\frac{-2\pi i} {\tau} H_{\text{CFT}}}|\gamma\rangle_{\text{Ishibashi}}
\end{split}
\end{equation} 
where the summation of $\gamma$ above is taken by applying the fusion rule of $\alpha\times \alpha$. In this sense, one can identify the spectrum of the Cardy states as that of the Ishibashi states (FIG. \ref{Schottky}). This identification of the chiral character and the amplitude of BCFT is called the Schottky double in the mathematical literature and the operator version is called the doubling trick in the BCFT literature\cite{Schweigert:2000ix}. For example, this identification and its application can be seen in \cite{Kawai:2002vd,Liu:2022gzl}. This explains the correspondence between the left-right entanglement and the topological entanglement of the topological ordered system which can be observed in a lot of fields in contemporary physics \cite{Das:2015oha,Qi_2012,Lou:2019heg,Wong:2017pdm,Hung:2019bnq,Zhang:2021djx,Cr_pel_2019}.

As a simple example, we consider the bosonic Ising BCFT which can be easily mapped to the fermionic BCFT. There exist three boundary conditions as labeled by the primary operators, $I$, $\epsilon$, $\sigma$, with conformal dimensions $h_{I}=0, h_{\epsilon}=1/2, h_{\sigma}=1/16$ as we have introduced in section \ref{Moore_Read}. Hence by computing the fusion rule, we can obtain the following Ishibashi states by applying the $S$ transformation above:
\begin{align}
|I\rangle_{\text{Cardy}} \langle I|_{\text{Cardy}} &\sim |I \rangle_{\text{Ishibashi}} \langle I |_{\text{Ishibashi}}, \\
|\epsilon\rangle_{\text{Cardy}}\langle\epsilon|_{\text{Cardy}} &\sim|I \rangle_{\text{Ishibashi}} \langle I |_{\text{Ishibashi}},
\end{align}
for $Z_{2}$ invariant parts, and
\begin{equation}
\begin{split}
&|\sigma\rangle_{\text{Cardy}} \langle\sigma|_{\text{Cardy}}\sim \left(|I \rangle_{\text{Ishibashi}}+e^{i\theta} |\epsilon \rangle_{\text{Ishibashi}}\right) \\
&\left(\langle I |_{\text{Ishibashi}}+e^{-i\theta} \langle\epsilon |_{\text{Ishibashi}}\right),
\end{split}
\end{equation}
for $Z_{2}$ noninvariant part, where $\theta$ is an arbitrary parameter. One may notice that the choice $\theta=0, \pi$ gives a diagonal basis. Hence it is natural to relate this choice to the zero modes of $\sigma$. One may also notice that the Cardy states of identity and $\epsilon$ fields result in the same Ishibashi state. This can be understood as the parity symmetry breaking. Moreover, it should also be interesting to note that one cannot obtain the diagonal sets of states from the right-hand side without restriction. We can do the same operation in principle with the simple current, but this depends on the choice of the Cardy states (and may depend on the definition of the conjugation).

\begin{figure}[htbp]
\begin{center}
\includegraphics[width=0.5\textwidth]{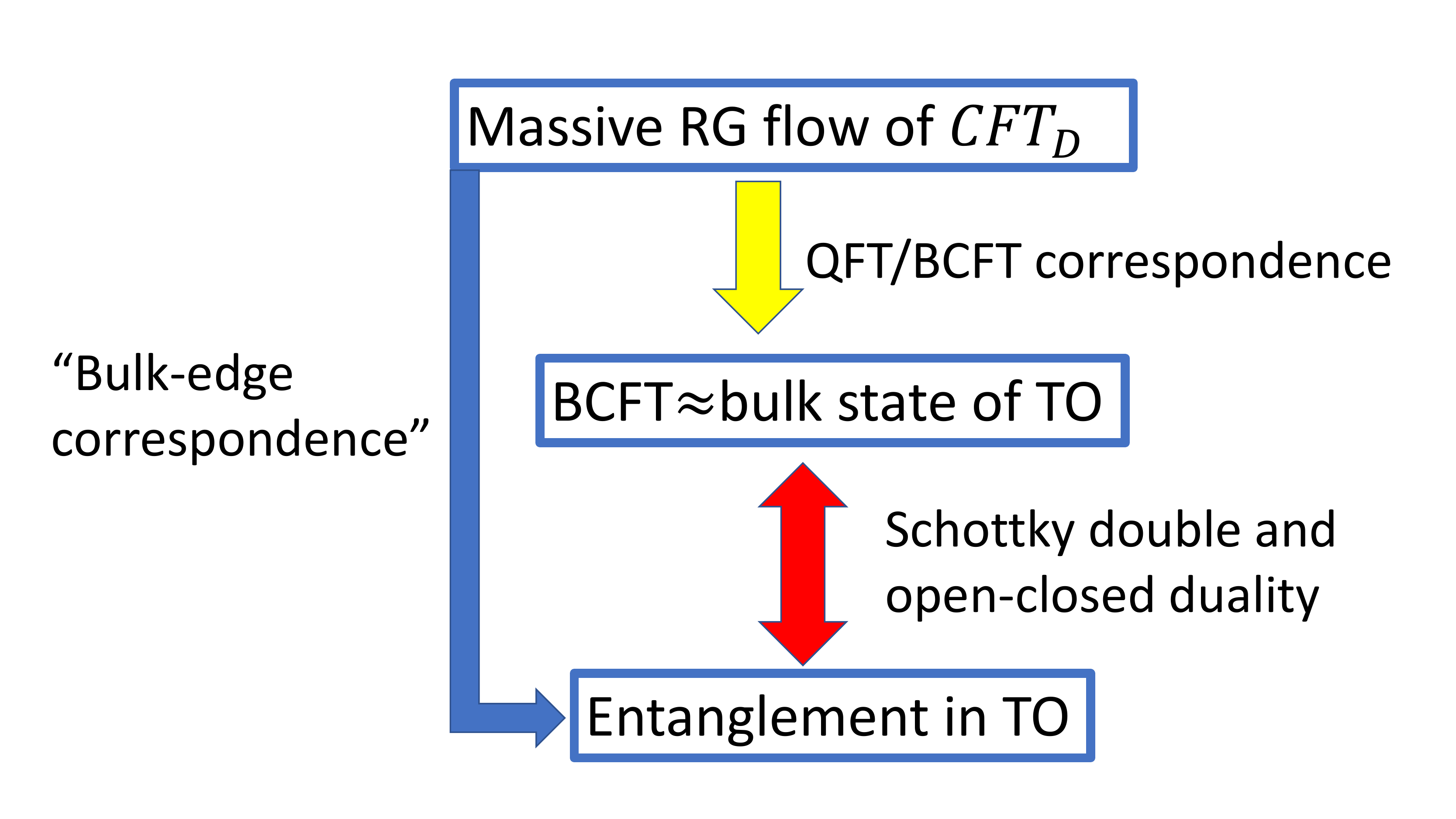}
\caption{A RG understanding of entanglement and bulk states in TO. In the existing literature, only the blue arrow has been proposed and studied commonly. We filled the gap by using the traditional argument by the QFT/BCFT correspondence combined with the handle decomposition of the manifold and open-closed duality of BCFT. It should be stressed that the meaning of the bulk-edge correspondence in our work is more fundamental than that in this figure and the literature.}
\label{bulk}
\end{center}
\end{figure}

One should also note that only the QFT/BCFT correspondence is assumed in this section, while the modular invariance is not necessary. Hence one can say that whereas the modular invariance gives restriction for gapping out the system under the bulk-edge correspondence, the existence of the BCFT (or non-negative integer matrix representation, mathematically) may ensure well-definedness of the topological degeneracy and topological entanglement entropy\cite{Fuchs:1994sq, BoyleSmith:2022duw, Gannon:2001ki} \footnote{As we have discussed for the protected edge modes, we expect the bulk NS sector or Lagrangian subalgebra becomes more stable after gapping the bulk R sector, but this needs further verifications.}. This construction of the bulk TO and the topological entanglement can be applied in higher dimensional systems with bulk-edge correspondence at least in principle, and one can consider this mapping together with the handle decomposition of the manifold \cite{Nakayama:2015mva,Chen_2016}. In short, one can say the appearance of the bulk topological degeneracy of $D+1$ dimensional system can be interpreted as a consequence of the RG flow of the $D$ dimensional CFT and the topological entanglement is representing this RG flow by the high-low temperature duality(FIG. \ref{bulk}).

\begin{figure}[htbp]
\begin{center}
\includegraphics[width=0.5\textwidth]{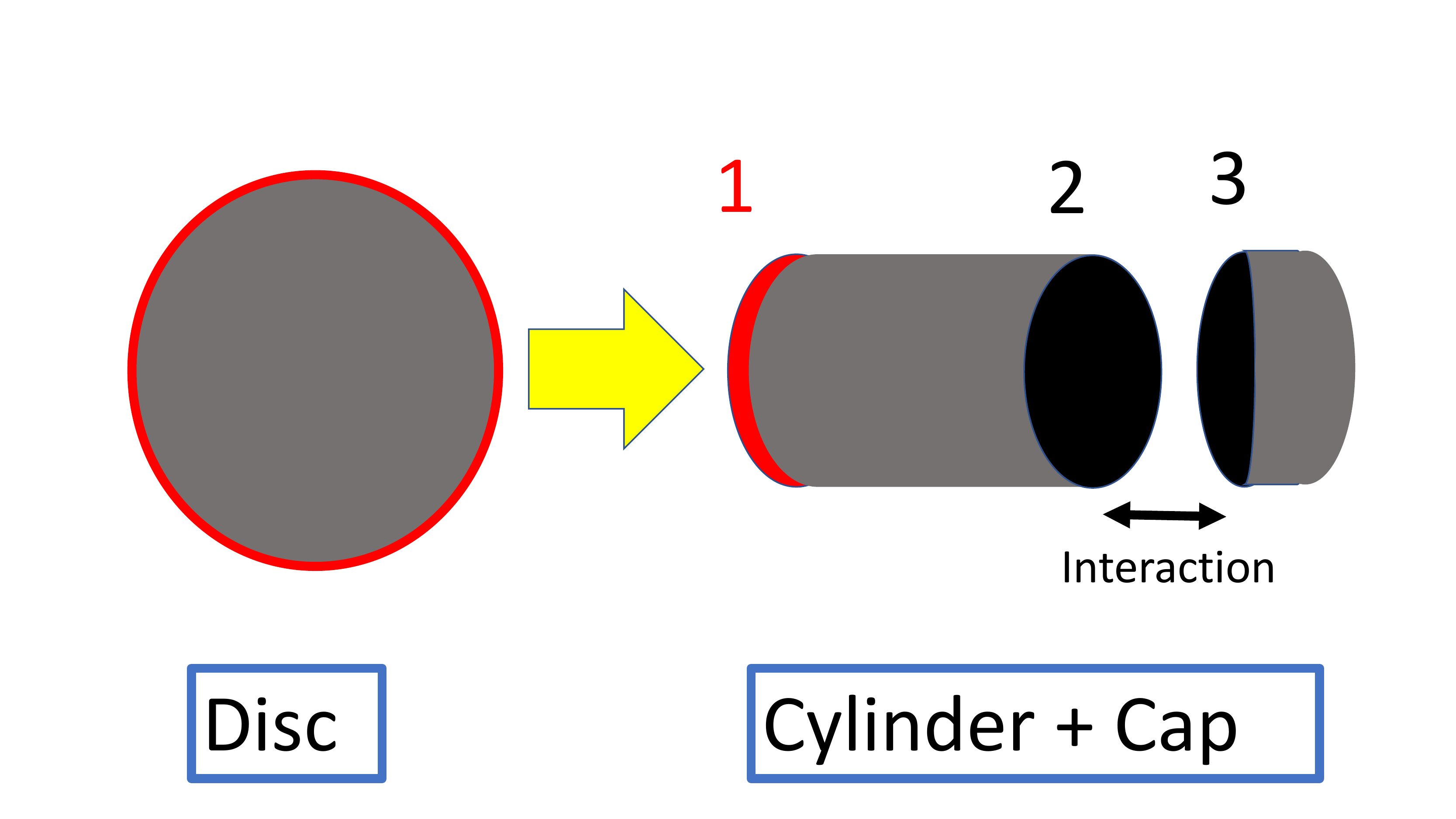}
\caption{Decomposition of a disc to a cylinder and cap }
\label{plane}
\end{center}
\end{figure}

Schematically, one can divide the plane as a cylinder with two edges labeled by the index ``$1$" and ``$2$", and a smaller plane or a cap labeled by ``$3$", see FIG. \ref{plane} (and figure 1 in \cite{Kobayashi:2022vgz}). The edge of the cap and the edge ``$2$" of the cylinder  interact each other. Naively, one can consider the following Hamiltonian:
\begin{equation}
H_{\text{cylinder},1}+H_{\text{cylinder},2}+H_{\text{cap},3}+\lambda H_{\text{int},2-3}
\end{equation}

By assuming all edge is constructed fermionically, and the edge $2$ and $3$ are gapped out and goes to fermionic BCFT, we can obtain the partition function corresponding to the chiral edge modes in the fermionic FQHE as:
\begin{equation} 
Z_{\text{plane},1}=\sum_{i,r\in\text{untwisted}}(\Xi_{i+,\frac{r}{q}}+\Xi_{i-,\frac{r}{q}}).
\end{equation}

This may explain why the partition function $\Xi_{\sigma}$ seems not to appear as the protected edge modes in the Pfaffian FQHE\cite{Milovanovic:1996nj}. Similarly, the vanishing of the characters of the twisted states may be valid in other fermionic FQHE and generally depends on the type of bulk states (or fermionic boundary states equivalently).

It should also be noted that the plane partition function of the topological ordered system can be obtained from the $1+1$ dimensional CFT in annulus geometry by applying the doubling trick\cite{Schweigert:2000ix}. In this sense, contrary to the common belief, there exists mapping from the chiral edge modes to $1+1$ dimensional (B)CFT. However, this map changes the geometry. This interpretation may give some geometrical intuition to consider the chiral edge modes in FQHE, because the structure of the conformal tower is preserved with some shifts of the pseudo-energy in this geometry\cite{Cappelli:1992kf}. A benefit of our argument is that one can test (and even modify) our analysis as a form of the partition function, which can be completely visible and treatable both numerically and analytically in principle.

\section{Conclusions and outlooks}
\label{conclusion}

In this work, we have shown a systematic construction of the cylinder partition function of the fermionic FQHE. Whereas this object is a starting point for the construction of the protected edge modes of FQHE in the established context of the CFT/TQFT correspondence, it contains a lot of information for the bulk states and protected edge modes in other geometries. Our work gives a new paradigm for the investigation of topological ordered systems, as a unification of modern aspects of theoretical physics such as bulk and boundary RG, various correspondence or duality between QFTs and the modular property of the systems. For the concluding remark, we note several open problems which are related to our construction.

First, we have assumed the existence of a $Z_{2}$ simple current and the corresponding wavefunctions, but the systematic construction of such wavefunctions may be computationally difficult. In other words, one has to consider multiple Dotsenko-Fateev integrals or the corresponding multi-parameter ordinary differential equations \cite{Estienne:2010as,Estienne:2011qk} and choose solutions which correspond to the situation by their monodromies. This is followed by considering the $Z_{2}$ fermionization \cite{Petkova:1988cy,Furlan:1989ra}. Hence it might be necessary to use some special properties of the correlation functions described by the products of the simple currents.

Next, in our construction, we have assumed that the central charge of the CFT at the boundary of the system is preserved under the imaginary gapping process. However, this may be problematic when considering the recent development of the bulk and boundary RG because the degree of freedom can be preserved or even increase under the bulk perturbation. A typical example is the appearance of the Majorana edge modes in the Kitaev chain. Hence in this sense, one has to consider the boundary version of the symmetry-enriched CFT or RG to the original CFT, from which the bulk wavefunctions are constructed. Therefore, the resulting central charge of the boundary of the system even can increase in principle. This may give a physical motivation to investigate CFTs with a large central charge which can potentially contain a lot of information about RG flow. We hope that this gives some clue to understand the recent controversies related to $\nu=5/2$ FQHE states \cite{Mross_2018,Wang2017TopologicalOF}, for example. We will study this problem related to the connectivity of RG and its application in the condensed matter in the forthcoming paper\cite{Fukusumi_2022_c}.   

\begin{figure}[htbp]
\begin{center}
\includegraphics[width=0.5\textwidth]{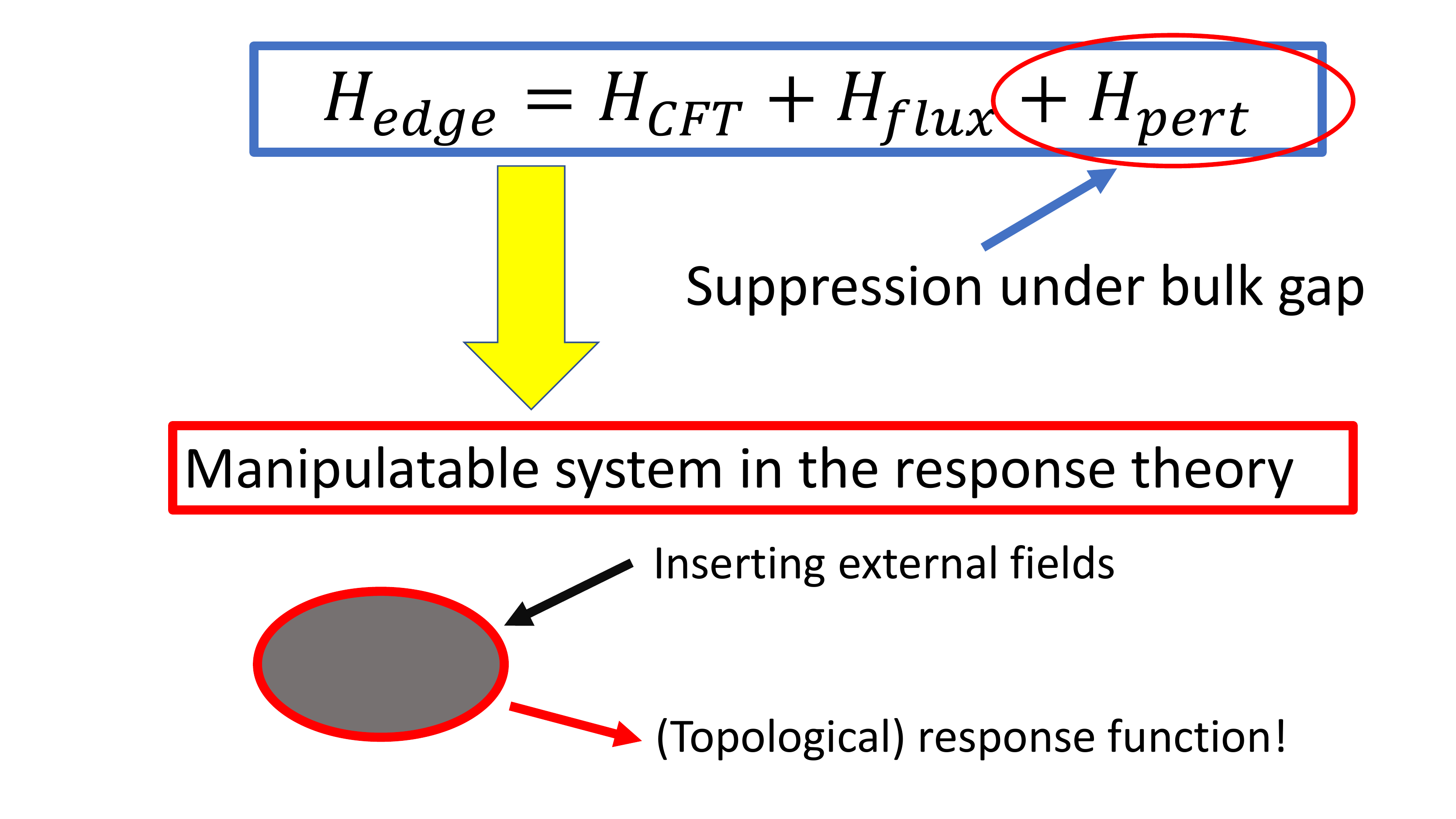}
\caption{A possible scenario for the emergence of response theory in the edge modes. $H_{flux}$ and $H_{pert}$ represent contributions from $U(1)$ flux (or external fields) and perturbations coming from the interaction in the lattice model. This realizability of response theory can be thought of as a result of emergent renormalizability in FIG. \ref{QFT} in Sec.\ref{sec:introduction}.}\label{response}
\end{center}
\end{figure}

Finally, we note a modern and general motivation in studying the protected edge modes of topological matters. This is related to the problem of controlling or predicting the properties of the gapless systems by using the response theory\cite{Kampen_1971}. As one of us has shown, it is difficult to apply the response theory in the Hamiltonian formalism to an interacting conductor\cite{Fukusumi:2021qwa,PhysRevB.104.205116,2022arXiv220809490F}\footnote{It may be worth noting that the difficulty of the response theory has been mentioned as a reason to introduce topological invariant in a gapped system in the work \cite{1985AnPhy.160..343K} by Kohmoto.}. From the RG perspective, this difficulty comes from the inappropriate ordering in treating the gauge fields (or the marginal perturbations) and the irrelevant interactions. In the response theory, small energy splitting caused by the irrelevant perturbation results in gigantic energy splittings from the finite flux and large system size. Hence, in the Hamiltonian formalism which should be valid for long wave limit and low temperature, the response theory contains subtleties.  To overcome this difficulty, one can take several approaches: by applying fine tuning to eliminate such dangerous irrelevant interactions \cite{PhysRevB.104.205116,2022arXiv220809490F,Urichuk_2021}, by considering Lagrangian formalism which corresponds to the high-temperature limit\cite{Lukyanov:1997wq,Prokof_ev_2000}, or by considering the response theory of the protected edge modes which may eliminate boundary irrelevant perturbations under the bulk gap. The third approach is speculative to some extent as we have discussed but it may be more testable and treatable in the experimental settings (FIG. \ref{response}). If this scenario succeeds, it may lead to a new possibility of manipulating the protected edge modes in predictable ways, e.g. serving as a first step for the universal topological quantum computation\cite{Mong_2014}.

\section{Acknowledgement}
YF thanks Yuji Tachikawa and Yunqin Zheng for the stimulating discussions, and Masaki Oshikawa for the discussion about the interpretation ambiguities of the modular invariance in the lattice model. The author thanks helpful comments from Juven Wang. He also thanks the organizers and speakers at the conference `Bootstat 2021: Conformal bootstrap and statistical models' at Orsay, France, for the introduction of a modern understanding of the operator formalism of CFT. This work is supported by the NTU grant for Nanyang Assistant Professorship and the National Research Foundation, Singapore under the NRF fellowship award (NRF-NRFF12-2020-005), and a Nanyang Technological University start-up grant (NTU-SUG). A large part of this work was accomplished in the conference `Topological interacting electron: in person', at Quy Nhon, Vietnam. 
\clearpage

\appendix

\section{Modular transformation for Riemann theta function}
In this section, we summarize the modular $S$ transformation property of the (eliptic) Riemann theta function which contains the plane partition function of the Laughlin wavefunction derived in \cite{Ino:1998by}. Here, let us introduce the following theta function with modular parameter $\tau$ as in the main text,
\begin{equation}
\theta^{+}_{\frac{r}{q}}(\tau)=\frac{1}{\eta (x)}\sum_{m=-\infty}^{\infty}x^{\frac{qm+r}{2q}}, x=e^{2\pi i \tau},
\end{equation}
where we have taken $r=0, 1, ...,q-1$ as an integer or half-integer, and $q$ as an positeve integer. As in the main text, we fix $r$ to an integer and denote $r+1/2$ as the half-integer case. Also, we introduce the following function,
\begin{equation}
\theta^{-}_{\frac{r}{q}}(\tau)=\frac{1}{\eta (x)}\sum_{m=-\infty}^{\infty}(-1)^{m}x^{\frac{qm+r}{2q}}, x=e^{2\pi i \tau}.
\end{equation} 
Then the modular $S$ transformation properties for the theta functions which are relevant in the main text are,
\begin{align}
\theta^{+}_{\frac{r}{q}}(-1/2\tau)&= \frac{1}{\sqrt{q}}\sum_{r'=0}^{q-1}e^{2\pi i rr'/q}\theta^{+}_{\frac{r'}{q}}(\tau),\\
\theta^{+}_{\frac{r+1/2}{q}}(-1/2\tau)&= \frac{1}{\sqrt{q}}\sum_{r'=0}^{q-1}e^{2\pi i (r+1/2)r'/q}\theta^{-}_{\frac{r'}{q}}(\tau),
\end{align}
By taking the parameters as corresponding to the situations in the main text, and considering the product of chiral and antichiral parts, one can obtain Eq. \eqref{modular_S_Laughlin} and \eqref{modular_S_Laughlin_2} in the main text.

\section{Bosonic partition function}
\label{bosonic_pf}
In this section, we see another type of partition function which does not have protected edge modes.
The bosonic modular $T$ invariant partition function which one can construct naively is,

\begin{equation}
\begin{split}
Z_{T-\text{inv}}^{\text{even}}&=\sum_{i, r}\left(\sum_{\delta=\pm}\Xi_{i\delta,\frac{r}{q}}^{\text{even}} \overline{\Xi}_{i\delta, \frac{r}{q}}^{\text{even}}+\Xi_{i\delta,\frac{r}{q}}^{\text{odd}} \overline{\Xi}_{i\delta, \frac{r}{q}}^{\text{odd}}\right) \\
&+\sum_{a, r}|\Xi_{a,\frac{r}{q}}|^{2},
\end{split}
\end{equation}
\begin{align}
Z_{T-\text{inv}}=\sum_{i, r}\left(|\Xi_{i+,\frac{r}{q}}|^{2}+|\Xi_{i-,\frac{r}{q}}|^{2}\right)+2\sum_{a, r}|\Xi_{a,\frac{r}{q}}|^{2},
\end{align}
where $Z_{T-\text{inv}}^{\text{odd}}$ can be obtained by applying the parity shift operation to $Z_{T-\text{inv}}^{\text{even}}$.
For $Z_{T-\text{inv}}$, one can see this is similar to the usual charge conjugate modular invariant partition function. However, the factor before $Z_{2}$ invariant fields are different from that of the charge-conjugated modular invariant. Hence only by gapping out the bosonic part simply, there exists a difficulty to obtain modular invariance. This is a consequence of our construction of the partition function by introducing the fermionic parity of the system. It should be noted that introducing the fermionic parity and half-flux quantum can lead to totally different structures of the partition function as we have explained in the main text.

\section{Parity of tricritical Ising model}

The tricritical Ising model has seven primary fields denoted as $\{ I, \epsilon , \epsilon' , \epsilon'' , \sigma, \sigma'\}$, with conformal dimensions  $\{ h_{I}=0, h_{\epsilon}=\frac{1}{10} , h_{\epsilon'}=\frac{3}{5} , h_{\epsilon''}=\frac{3}{2} , h_{\sigma}=\frac{3}{80}, h_{\sigma'}=\frac{7}{16}\}$. As can be seen from the conformal dimension, the operator $\epsilon''$
is the $Z_{2}$ simple current. (For a review and notations, see the discussions and reference in \cite{Balaska:2006yn})

The fusion rule is,
\begin{align}
\epsilon'' \times \epsilon''&=I , \\
\epsilon''  \times \epsilon&=\epsilon', \\
\epsilon'' \times \epsilon'&=\epsilon, \\
\epsilon \times \epsilon'&=\epsilon +\epsilon'' ,\\
\epsilon\times \epsilon&=\epsilon' \times \epsilon' =I+\epsilon', \\
\epsilon'' \times\sigma &=\sigma, \\
\epsilon \times \sigma &=\epsilon' \times \sigma =\sigma +\sigma', \\
\sigma \times \sigma &=I+\epsilon +\epsilon' +\epsilon'' ,\\
\epsilon'' \times \sigma'&=\sigma', \\
\epsilon\times\sigma'&=\epsilon'\times\sigma'=\sigma, \\
\sigma \times \sigma'&=\epsilon+\epsilon', \\
\sigma' \times \sigma' &=I+\epsilon''
\end{align}
As can be seen in the fusion rule, the parity of the model can be take as,
\begin{align}
\text{Parity even} &; I, \epsilon', \\
\text{Parity odd} &; \epsilon, \epsilon'', \\
\text{Zero mode} &; \sigma=\frac{e+m}{\sqrt{2}}, \sigma'=\frac{e'+m'}{\sqrt{2}}.
\end{align}
where we have introduced the semion basis $\sigma^{\text{even}}=e$, $\sigma^{\text{odd}}=m$, $\sigma'^{\text{even}}=e'$, $\sigma'^{\text{odd}}=m'$ as in the main text.
It should be noted that whereas BCFT and its boundary states give the following correspondence, $I\rightarrow +$, $\epsilon''\rightarrow -$, $\epsilon\rightarrow d+$, $\epsilon'\rightarrow d-$ where $\pm$ corresponds to spin fixed boundary conditions and $d\pm$ corresponds to the disordered boundary conditions in the $2-$dimesional statistical model. Hence the correspondence between the boundary condition of the spin in the lattice model and the fermionic parity of the operator is changed when there exists fugacity. By definition, one can see the set $\{ I, \epsilon'', e', m' \}$ satisfies the same fusion rule as that of the Ising model and forms the Tambara-Yamagami category.

In this basis, the fusion rule is,
\begin{align}
\epsilon'' \times \epsilon''&=I , \\
\epsilon''  \times \epsilon&=\epsilon', \\
\epsilon'' \times \epsilon'&=\epsilon, \\
\epsilon \times \epsilon'&=\epsilon +\epsilon'' ,\\
\epsilon\times \epsilon&=\epsilon' \times \epsilon' =I+\epsilon', \\
\epsilon'' \times e &=m, \\
\epsilon'' \times m &=e, \\
\epsilon \times e &=\epsilon' \times m=m+m', \\
\epsilon \times m &=\epsilon' \times e=e+e', \\
e \times e &=m\times m=I+\epsilon' ,\\
m\times e &=\epsilon+\epsilon'', \\
\epsilon''\times e'&=m', \\
\epsilon'' \times m'&=e', \\
\epsilon\times e'&=\epsilon' \times m'=m, \\
\epsilon\times m'&=\epsilon' \times e'=e, \\
e\times e'&= m\times m'=\epsilon', \\
e\times m'&= m\times e'=\epsilon, \\
e' \times e' &=m'\times m'=I, \\
e' \times m' &=\epsilon''.
\end{align}
Because of the fusion rule, one can see the fields $\epsilon$ and $\epsilon'$ do not have zero modes when assuming the fusion matrix should be represented as an integer matrix. Our results are consistent with the recent mathematical work by assigning $8$ gauge choices as $\sigma=\frac{e\pm m}{\sqrt{2}}, \sigma'=\frac{\pm e'\pm m'}{\sqrt{2}}$ \cite{Chang:2022hud}.

\bibliography{Cardy_v2}

\end{document}